\documentclass{ws-procs975x65}

\begin{document}

\title{Theoretical interpretation of luminosity and spectral properties of GRB 031203}

\author{C.L. Bianco,$^{1,2}$ M.G. Bernardini,$^{1,2}$ P. Chardonnet,$^{1,4}$ F. Fraschetti,$^5$ R. Ruffini,$^{1,2,3}$ S.-S. Xue$^{1}$}

\address{
$^1$ ICRANet and ICRA, Piazzale della Repubblica 10, I-65122 Pescara, Italy.\\
$^2$ Dip. di Fisica, Universit\`a di Roma ``La Sapienza'', Piazzale Aldo Moro 5, I-00185 Roma, Italy.\\
$^3$ ICRANet, Universit\'e de Nice Sophia Antipolis, Grand Ch\^ateau, BP 2135, 28, avenue de Valrose, 06103 NICE CEDEX 2, France.\\
$^4$ Universit\'e de Savoie, LAPTH - LAPP, BP 110, F-74941 Annecy-le-Vieux Cedex, France.\\
$^5$ Laboratoire AIM, CEA/DSM - CNRS - Universit\'e Paris Diderot, Irfu/Service d'Astrophysique, F-91191 Gif sur Yvette C\'edex, France.\\
E-mails: bianco@icra.it, maria.bernardini@icra.it, chardon@lapp.in2p3.fr, fraschetti@icra.it, ruffini@icra.it, xue@icra.it.
}

\begin{abstract}
We show how an emission endowed with an instantaneous thermal spectrum in the co-moving frame of the expanding fireshell can reproduce the time-integrated GRB observed non-thermal spectrum. An explicit example in the case of GRB 031203 is presented.
\end{abstract}

\bodymatter

\section{Introduction}

One aim of our model (see e.g. Ref.~\refcite{2007AIPC..910...55R} and references therein) is to derive from first principles both the luminosity in selected energy bands and the time resolved/integrated spectra of GRBs.\cite{2004IJMPD..13..843R} The luminosity in selected energy bands is evaluated integrating over the equitemporal surfaces (EQTSs)\cite{2004ApJ...605L...1B,2005ApJ...620L..23B} the energy density released in the interaction of the optically thin fireshell with the CircumBurst Medium (CBM) measured in the co-moving frame, duly boosted in the observer frame. The radiation viewed in the co-moving frame of the accelerated baryonic matter is assumed to have a thermal spectrum and to be produced by the interaction of the CBM with the front of the expanding baryonic shell.\cite{2004IJMPD..13..843R}

\section{The instantaneous GRB spectra}

In Ref.~\refcite{2005ApJ...634L..29B} it is shown that, although the instantaneous spectrum in the co-moving frame of the optically thin fireshell is thermal, the shape of the final instantaneous spectrum in the laboratory frame is non-thermal. In fact, as explained in Ref.~\refcite{2004IJMPD..13..843R}, the temperature of the fireshell is evolving with the co-moving time and, therefore, each single instantaneous spectrum is the result of an integration of hundreds of thermal spectra with different temperature over the corresponding EQTS. This calculation produces a non thermal instantaneous spectrum in the observer frame.\cite{2005ApJ...634L..29B} Another distinguishing feature of the GRBs spectra which is also present in these instantaneous spectra is the hard to soft transition during the evolution of the event.\cite{1997ApJ...479L..39C,1999PhR...314..575P,2000ApJS..127...59F,2002A&A...393..409G} In fact the peak of the energy distributions $E_p$ drift monotonically to softer frequencies with time.\cite{2005ApJ...634L..29B} This feature explains the change in the power-law low energy spectral index\cite{1993ApJ...413..281B} $\alpha$ which at the beginning of the prompt emission of the burst ($t_a^d=2$ s) is $\alpha=0.75$, and progressively decreases for later times.\cite{2005ApJ...634L..29B} In this way the link between $E_p$ and $\alpha$ identified in Ref.~\refcite{1997ApJ...479L..39C} is explicitly shown.

\section{The time-integrated GRB spectra - Application to GRB 031203}

The time-integrated observed GRB spectra show a clear power-law behavior. Within a different framework (see e.g. Ref.~\refcite{1983ASPRv...2..189P} and references therein) it has been argued that it is possible to obtain such power-law spectra from a convolution of many non power-law instantaneous spectra monotonically evolving in time. This result was recalled and applied to GRBs\cite{1999ARep...43..739B} assuming for the instantaneous spectra a thermal shape with a temperature changing with time. It was shown that the integration of such energy distributions over the observation time gives a typical power-law shape possibly consistent with GRB spectra.

Our specific quantitative model is more complicated than the one considered in Ref.~\refcite{1999ARep...43..739B}: the instantaneous spectrum here is not a black body. Each instantaneous spectrum is obtained by an integration over the corresponding EQTS: \cite{2004ApJ...605L...1B,2005ApJ...620L..23B} it is itself a convolution, weighted by appropriate Lorentz and Doppler factors, of $\sim 10^6$ thermal spectra with variable temperature. Therefore, the time-integrated spectra are not plain convolutions of thermal spectra: they are convolutions of convolutions of thermal spectra.\cite{2004IJMPD..13..843R,2005ApJ...634L..29B}

\begin{figure}
\includegraphics[width=\hsize,clip]{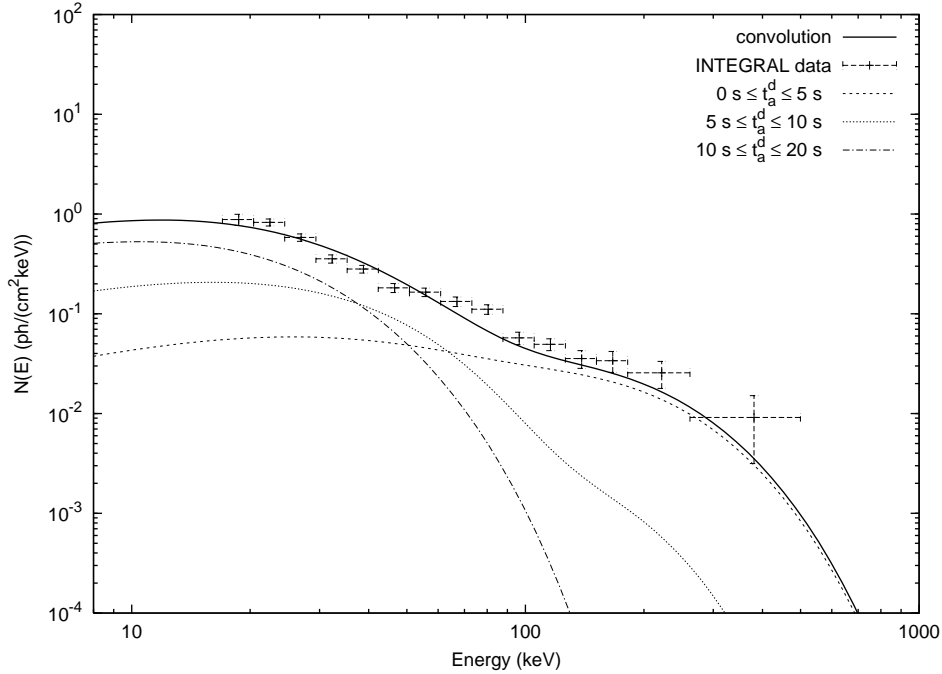}
\caption{Three theoretically predicted time-integrated photon number spectra $N(E)$, computed for GRB 031203,\cite{2005ApJ...634L..29B} are here represented for $0 \le t_a^d \le 5$ s, $5 \le t_a^d \le 10$ s and $10 \le t_a^d \le 20$ s (dashed and dotted curves), where $t_a^d$ is the photon arrival time at the detector.\cite{2001ApJ...555L.107R,2005ApJ...634L..29B} The hard to soft behavior is confirmed. Moreover, the theoretically predicted time-integrated photon number spectrum $N(E)$ corresponding to the first $20$ s of the ``prompt emission'' (black bold curve) is compared with the data observed by INTEGRAL\cite{2004Natur.430..646S}. This curve is obtained as a convolution of 108 instantaneous spectra, which are enough to get a good agreement with the observed data. Details in Ref.~\refcite{2005ApJ...634L..29B}.}
\label{031203_spettro}
\end{figure}

In Fig.~\ref{031203_spettro} we present the photon number spectrum $N(E)$ time-integrated over the $20$ s of the whole duration of the prompt event of GRB 031203 observed by INTEGRAL:\cite{2004Natur.430..646S} in this way we obtain a typical non-thermal power-law spectrum which results to be in good agreement with the INTEGRAL data\cite{2004Natur.430..646S,2005ApJ...634L..29B} and gives a clear evidence of the possibility that the observed GRBs spectra are originated from a thermal emission.\cite{2005ApJ...634L..29B}

\end{document}